\documentclass[12pt]{article}
\usepackage{graphicx}
\newsavebox{\sboxpubnumber}
\newsavebox{\sboxpubdate}
\newcommand{\pubdate}[1]{\begin{lrbox}{\sboxpubdate}{#1}\end{lrbox}}

\newcommand{\Title}[1]{\begin{center} {\Large #1 } \end{center}}
\newcommand{\Author}[1]{\begin{center}{ \sc #1} \end{center}}
\newcommand{\Address}[1]{\begin{center}{ \it #1} \end{center}}

\newenvironment{Abstract}{\begin{quotation}  }{\end{quotation}}
\newenvironment{Presented}{\begin{quotation} \begin{center}
             PRESENTED AT\end{center}\bigskip
      \begin{center}\begin{large}}{\end{large}\end{center}
      \end{quotation}}
\newcommand{\Acknowledgements}{\bigskip  \bigskip \begin{center} \begin{large}
             \bf ACKNOWLEDGEMENTS \end{large}\end{center}}
\newcommand {\lsim}{\mbox{$\:\stackrel{<}{_{\sim}}\:$} }

\def\be{\begin{equation}}
\def\ee{\end{equation}}
\def\bea{\begin{eqnarray}}
\def\eea{\end{eqnarray}}
\def\om{\Omega_{\rm m}}
\def\ol{\Omega_{\Lambda}}
\def\ob{\Omega_{\rm b}}
\def\obh2{\Omega_{\rm b}h^{2}}
\def\og{\Omega_{\gamma}}
\def\on{\Omega_{\nu}}
\def\ok{\Omega_{k}}
\def\ocrit{\Omega_{crit}}
\def\otot{\Omega_{tot}}
\def\oc{\Omega_{\rm c}}

\def\orel{\Omega_{\rm rel}}
\def\t{t_{\rm o}}
\def\beq{\begin{equation}}
\def\eeq#1{\label{#1}\end{equation}}
\def\eeqn{\end{equation}}

\def\beqa{\begin{eqnarray}}
\def\eeqa#1{\label{#1}\end{eqnarray}}
\def\eeqan{\end{eqnarray}}

\let\bar=\overbar

\def\Dslash{\not{\hbox{\kern-4pt $D$}}}
\def\dslash{\not{\hbox{\kern-2pt $\del$}}}

\def\ee{e^+e^-}

\def\msb{{\bar{\ssstyle M \kern -1pt S}}}

\begin{document}
\begin{titlepage}
\pubdate{11th December 2001}                
%\pubnumber{~} %preprint number(s)
\vfill
\Title{Cosmological Parameters}
\vfill
\Author{Charles H. Lineweaver}
\Address{School of Physics, University of New South Wales, Sydney, Australia}
\vfill
\begin{Abstract}
 This article briefly summarizes the increasingly precise observational 
estimates of the cosmological parameters. After 
three years on the stump, the $\Lambda$CDM model is still the leading candidate.
Although the Universe is expanding, our picture of it is coming
together. 
\end{Abstract}
\vfill
\begin{Presented}
    COSMO-01 \\
    Rovaniemi, Finland, \\
    August 29 -- September 4, 2001
\end{Presented}
\vfill
\end{titlepage}
\def\thefootnote{\fnsymbol{footnote}}
\setcounter{footnote}{0}
%%%%%%%%%%%%%%%%%%%%%%%%%%%%%%%%%%%%%%%%%%%%%%%%%%%%%
\section{Cosmic Connections, Complementarity, Concordance and Consistency}
\begin{figure}[!ht]
    \centering
    \resizebox{8cm}{8cm}{\includegraphics[60,80][538,608]{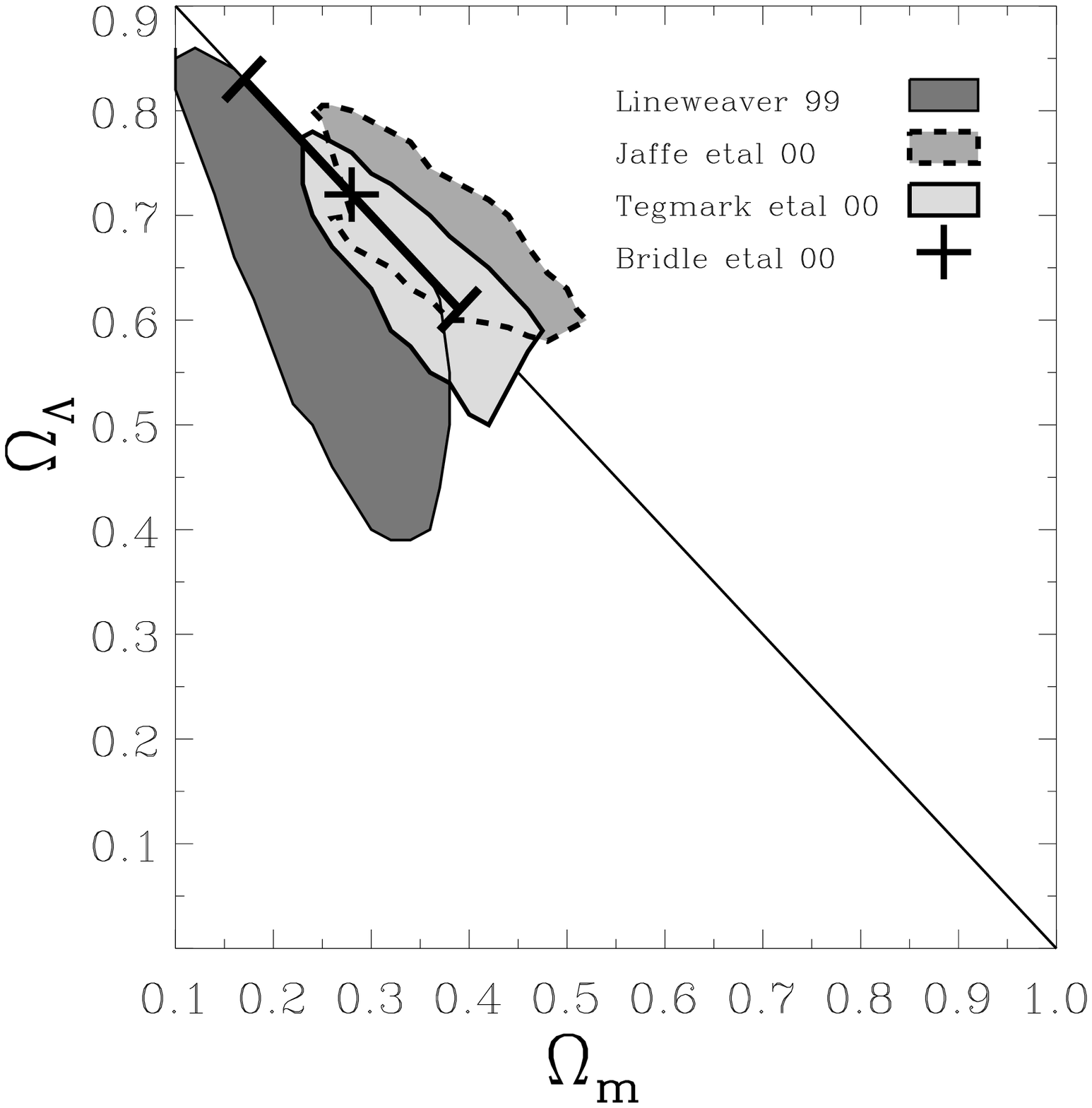}}
%    \resizebox{8cm}{8cm}{\includegraphics[-65,250][400,671]{energydensitiesSm.eps}}
    \resizebox{8cm}{8cm}{\includegraphics{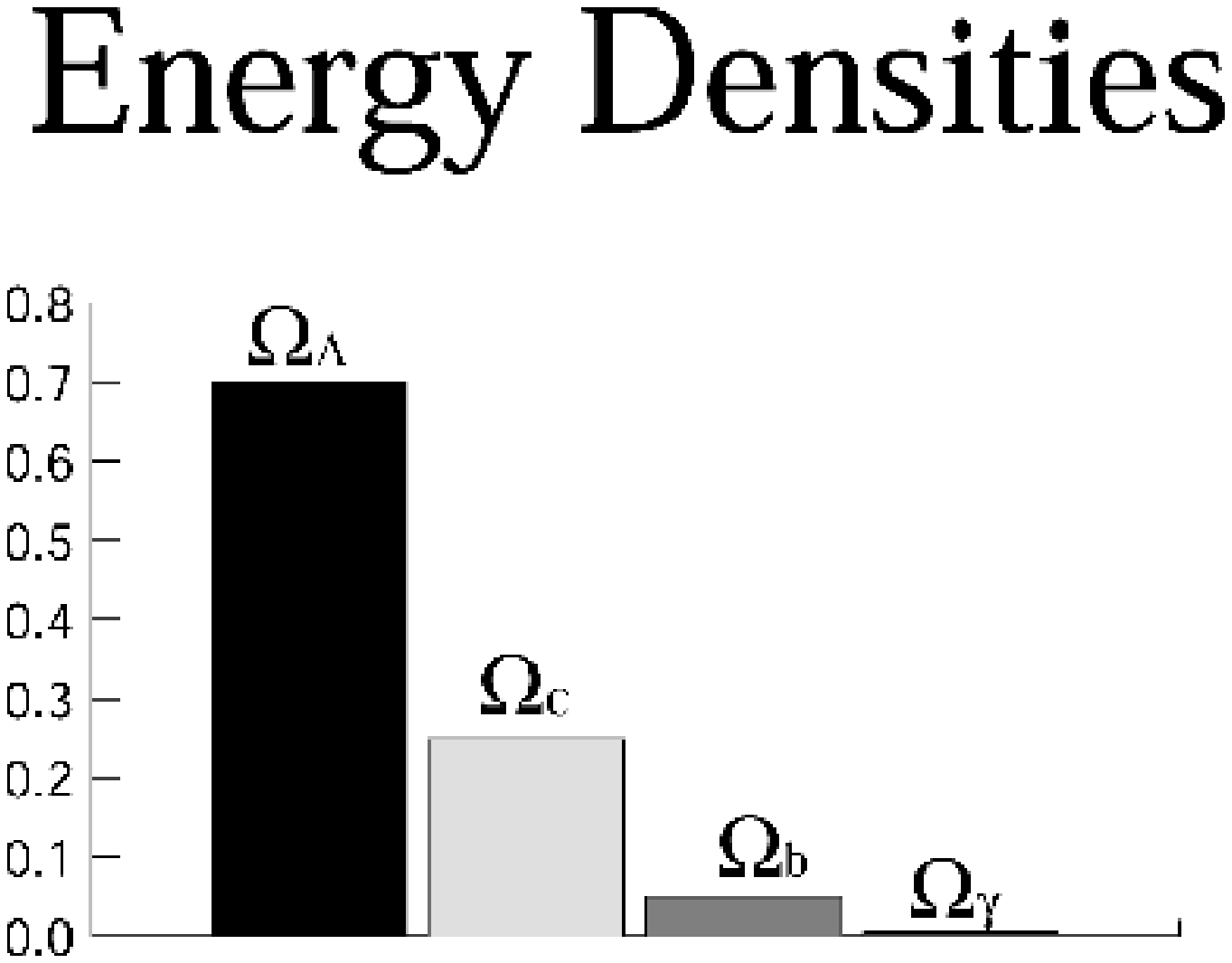}}
%    \resizebox{6cm}{6cm}{\includegraphics{energydensitiesnew.ps}}
    \caption{Various combinations of cosmic microwave background (CMB), 
supernovae and other 
    observational constraints favor the region $(\om, \ol) \approx (0.3, 0.7)$
    \cite{L98} --
% \cite{Jaffe} \cite{TZH} \cite{Bridle} \cite{sn}
\cite{Roos}.
    The most recent analyses, \cite{Stompor} \cite{deBernardis},
    continue to favor this region.
    The composition of the Universe is thus, energy of the vacuum: 
     $70\% \pm 10\%$, matter: $30\% \pm 10\%$ (cold dark matter: 
     $26\% \pm 10\%$, normal baryonic matter: $4\% \pm 1\%$), with 
     negligible energy density from photons.
%About 13.4 billion years ago the energy densities of photons and relativistic 
%matter played an important role in the dynamics of the Universe but that 
%is no longer the case.
      The neutrino energy density is poorly constrained and may be
      as large as the baryonic energy density. About 13\% of the matter 
     in the Universe is 
     baryonic ($\frac{\ob}{\om} = \frac{0.04}{0.3} = 0.13$). %^{+0.12}_{-0.05}$).
     The baryons can be further divided into $3\%$ warm invisible gas, 
     $0.5\%$ optically visible stars and $0.5\%$ hot gas visible in the 
     x-rays \cite{Fukugita}.  The $\om$ in the top plot is equal to the 
     sum of the $\oc$ and $\ob$ in the lower plot.}
    \label{f:omol}
\end{figure}

If your model of the Universe is a turtle, you want to know how big 
the turtle is, how old the turtle is, where the turtle came from and, in some
obscure animistic models, how fast the turtle is expanding.
Cosmological parameters are the observable quantities that most cosmologists
think are important. In the context of general relativity and the 
hot big bang model, cosmological parameters are the numbers that, when
inserted into the Friedmann equation, 
\begin{equation}
H^{2} = H_{o}^{2}\left[\ol + \ok \;a^{-2} + \om \;a^{-3} + \orel \;a^{-4}\right],
\label{eq:Friedmann}
\end{equation}
%(Eq. 1)  % \ref{eq:Friedmann}) 
best describe our particular observable Universe.
The expansion is parametrized by 
Hubble's constant, $H_{o} = \dot{a}/a$, where $a$ is the scale 
factor of the Universe.
Observational estimates of the parameters in this equation, 
$H_{o}$, $\ol$, $\ok$, $\om$ and $\orel$ (and their subcomponents)
have been derived from hundreds of observations and analyses
(e.g. Fig.~\ref{f:omol}). Table 1 is my attempt to summarize this
immense body of work.

%%%%%%%%%%%%%%%%%%%%%%%%%%%%%%%%%%%%%%%%%%%%%%%%%%%%%%%%%%%
{\footnotesize    %\small    %\scriptsize
\begin{table}[!hb]
\begin{center} 
\caption{Cosmological Parameters (Background)}  %\\
\begin{tabular}{|l|l|l|l|} \hline
\multicolumn{1}{|c}{Parameter} &
\multicolumn{1}{|c}{Estimate} &
\multicolumn{1}{|c}{Sub-Components} &
\multicolumn{1}{|c|}{References}\\
\hline                          %\tableline
%                                  & Main Component & Sub-Component    &\\
%\hline
%Parameter                     & Component               &Sub-Component   & References\\
cosmological constant$^{a}$       &$\ol = 0.7 \pm 0.1   $& &\cite{L98} --
%\cite{L99}\cite{Jaffe}\cite{TZH}\cite{Bridle}\cite{Stompor}\cite{sn}\cite{Roos}
\cite{deBernardis}\\
matter$^{b}$                &$\om = 0.3 \pm 0.1   $& &  ` ' + \cite{om}\\
\hfill cold dark matter            &    &$\oc= 0.26 \pm 0.1     $& `  ' + \cite{bbn}\\
\hfill baryonic matter$^{c}$       &    &$\ob  = 0.04 \pm 0.01  $&\cite{bbn}\\ 
relativistic component$^{d}$        &$ 0.01 \lsim \orel  \lsim 0.05$&  &\cite{Wang} \cite{Durrer} \cite{nu}\\
\hfill       neutrinos$^{e}$       &    &$ 0.01 \lsim \on  \lsim 0.05$& `  '\\
\hfill       photons$^{f}$         &    &$\og=4.8^{+1.3}_{-0.9} \times 10^{-5}$&\cite{tcmb}\\
%\hline
Hubble's constant$^{g}$       &$h= 0.72 \pm 0.08   $& &\cite{h}\\
age of Universe$^{h}$         &$\t = 13.4  \pm 1.6$ Gyr & &\cite{L99}\cite{Jaffe}\cite{TZH}\cite{Ferreras}\\
geometry$^{i}$                &$\ok = 0.00 \pm 0.06 $& &` ' + \cite{Wang}\cite{deBernardisNat}\cite{Dodelson}\\
equation of state$^{j}$          &$w = -1.0^{+0.4}          $& &\cite{w}  \\
deceleration parameter$^{k}$     &$q_{o} = -0.05 \pm 0.15 $& &\cite{sn}\\ 
CMB temperature                  &$T_{\rm CMB} = 2.725 \pm 0.001$ K& &\cite{tcmb}\\
%\tableline 
\hline
\end{tabular}\\
\end{center}
\scriptsize    %\small    %\tiny
\noindent
$^{a}$ $\ol = \frac{\rho_{\Lambda}}{\rho_{crit}}$, $\rho_{crit} = \frac{3\: H_{o}^{2}}{8 \pi G}$, 
$\rho_{\Lambda} = \frac{\Lambda}{3\;H_{o}^{2}}$,
$^{b}$ $\om = \oc + \ob$,  
$^{c}$ $\obh2=0.020\pm 0.002$ \cite{bbn}, 
$^{d}$ $\orel = \on +\og$, 
$^{e}$ $ 0.04\; eV < m_{\nu,\tau} < 4.4\; eV$ \cite{Wang},\cite{Durrer},
$^{f}$ $\og =2.47 \times 10^{-5}\; h^{-2}\; T_{2.725}^{4}$ \cite{Scott},
$^{g}$ $ h = H_{o}/100 \; km^{-1} s^{-1}Mpc^{-1}$,
$^{h}$ $\t = h^{-1}f(\om, \ol)$, see Fig.\ref{f:age},
$^{i}$ $\ok = 1 - \otot$,  $\otot = \ol + \om + \orel$, $\ok = 0$ (flat), $>0$ (open), $<0$ (closed), thus
       $\otot = 1.00 \pm 0.06$,
$^{j}$ $p = w\rho$,
$^{k}$ $q_{o} = \om - \ol/2$
\end{table}
}   %end scriptsize
%\normalsize
%%%%%%%%%%%%%%%%%%%%%%%%%%%%%%%%%%%%%%%%%%%%%%%%%%%%%%%%%%
%0.3,0.7 with h = 72 - 13.2
%                 70 - 13.6
%     therefore   71 - 13.4

\clearpage
%%%%%%%%%%%%%%%%%%%%%%%%%%%%%%%%%%%%%%%%%%%%%%%%%%%%
%%%%%%%%%%%%%%%%%%%%%%%%%%%%%%%%%%%%%%%%%%%%%%%%%%%%
%%%%%%%%%%%%%%%%%%%%%%%%%%%%%%%%%%%%%%%%%%%%%%%%%%%%
%\begin{figure}[p]
%    \centering
%    \includegraphics[height=9cm]{energydensities.eps}
%    \caption{The composition of the Universe. 
%Energy of the vacuum: $70\% \pm 10\%$, 
%cold dark matter: $26\% \pm 10\%$, normal baryonic matter: $4\% \pm 1\%$,
%with negligible energy density from photons
% and neutrinos although the neutrino energy density is poorly constrained and may be
% as large as the baryonic energy density.
%    \label{f:energydensities}
%\end{figure}
%%%%%%%%%%%%%%%%%%%%%%%%%%%%%%%%%%%%%%%%%%%%%%%%%%%%
\clearpage

Various methods to extract cosmological parameters from cosmic microwave
background (CMB) and non-CMB 
observations are forming an ever-tightening network of interlocking constraints.
CMB observations tightly constrain  $\ok$, while type Ia supernovae 
observations tightly constrain $q_{o}$. Since lines of contant $\ok$ and 
constant $q_{o}$ are nearly orthogonal in the $\om-\ol$ plane, 
combining these measurements optimally constrains our Universe 
to a small region (Fig. 1).
Four years ago when $\ol$ was assumed to be zero, the critical density,
$\ocrit = \frac{3 H_{o}^{2}}{8\pi G}$ {\em was} 
critical -- it determined the fate of the universe -- whether it would expand
forever or recollapse. Currently, the notion of critical density has lost 
much of its importance. That role has been usurped by $\ol$; 
if $\ol > 0$ the universe will expand forever.

The upper limit on the energy density of neutrinos comes from the shape of the
small scale power spectrum. If neutrinos make a significant contribution
to the density, they suppress the growth of small scale structure
by free-streaming out of over-densities.
The CMB power spectrum is not sensitive to such suppression and is not a good way to
constrain $\on$.
Hubble scholars used to be irreconcilably divided into camps described by
a bimodal distribution peaking at $H_{o} = 50$  and  $H_{o} = 90$. 
These peaks seem to have merged into a more agreeable Gaussian distribution
peaking between 65 and 80  with error bars from
hostile groups now overlapping. 

The parameters in Table 1 are not independent of each other. 
The elongated contours in the top plot of Fig. 1 is one example
of correlation.
Another example is the age of the Universe,  $\t = h^{-1}f(\om, \ol)$.
Estimates of $h, \ol, \om$ can be inserted into  Eq.~\ref{eq:Friedmann}. 
Integration then yields the age of the Universe 
($\orel$ is negligible and $\ok = 1- \om - \ol \approx 0$). 
If the Universe is to make sense, independent
determinations of $\ol$, $\om$ and $h$ and the minimum age of the Universe
must be consistent with each other.
This is now the case (Fig. 2).
Presumably we live in a Universe 
which corresponds to a single point in 
multidimensional parameter space.
Estimates of $h$ from HST Cepheids
and the CMB must overlap. Deuterium and CMB determinations of $\obh2$
should be consistent. Regions of the $\om - \ol$ plane favored by supernovae
and CMB must overlap with each other and with other independent constraints.
This is the case \cite{L99}.

The proportionality constant $w$ in the equation of state, $p = w\rho$, is 
important in deciding whether the generalization of $\ol$ 
into a time varying $\ol$ (i.e. quintessence) is necessary.  So far 
the observations seem to be favoring the simplest case $w= -1$, 
(i.e. pure cosmological constant) and do not call for this generalization.

Just as $h$ describes the first derivative of the scale factor, the 
deceleration parameter $q_{o}$ describes the 
second derivative. The redshifts and apparent magnitudes of type Ia 
supernovae have been used to find that $q_{o} \lsim 0$ -- the expansion of the Universe
is accelerating.
The geometry of the Universe does not seem to 
be like the surface of a ball ($\ok < 0$) nor like a saddle ($\ok > 0$) 
but seems to be flat ($\ok \approx 0$) to the precision of our 
current observations.

\clearpage
%%%%%%%%%%%%%%%%%%%%%%%%%%%%%%%%%%%%%%%%%%%%%%%%%%%%
\begin{figure}[!h]
    \centering
    \includegraphics[height=15cm,width=14cm]{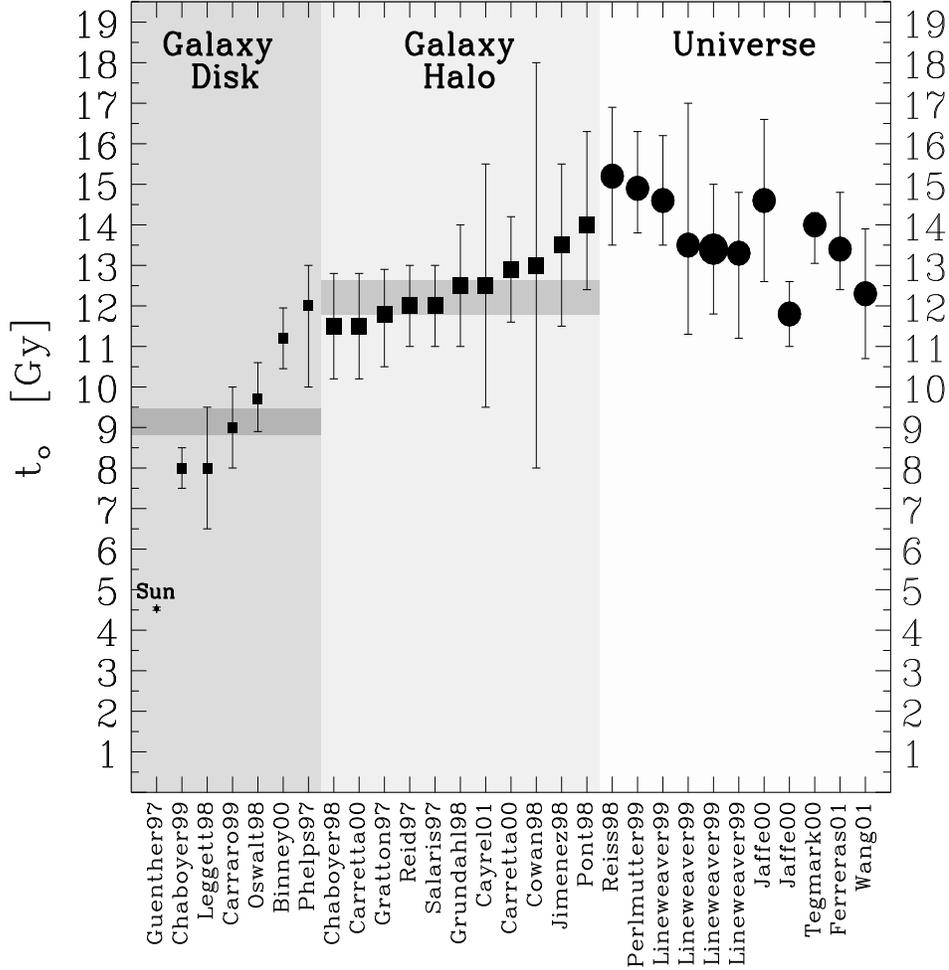}
    \caption{Estimates of the age of the Universe based on Eq. 1 are plotted
on the far right (`Universe') and can be compared to lower limits on the age of
the Universe from age estimates of the halo and disk of our Milky Way Galaxy.
Many different techniques, data sets and analysis methods were used to obtain
these estimates. Figure modified from \cite{L99}.}
    \label{f:age}
\end{figure}
%%%%%%%%%%%%%%%%%%%%%%%%%%%%%%%%%%%%%%%%%%%%%%%%%%%%

%Finally the average temperature of the CMB is
%a parameter used in many ways. notably, its time-dependence has been 
%convincingly measured and provides further support to the hot big bang model.

%Optical depth to rescattering $\tau < 0.2$ at 95\% CL. \cite{Wang}.
%Percival's version of concordance in the $\om h - \ob/\om$ plane
%of CMB, BBN and Clusters.
 
%There didn't use to be talks on `cosmological parameters'.
%Allan Sandage said there are only 2 parameters, $H_{o}$ and $q_{o}$.
%ie the expansion of the universe and the acceleration of the universe
%$a(t)$. 

\section{Background and the Bumps on it}

%General Relativity plus isotropy plus homogeneity 
%yield the hot big bang description of the Universe,
%the Friedmann-Robertson-Walker metric and the 
%Friedman Equation \cite{Kelvin}.

Equation \ref{eq:Friedmann} is our hot big bang description 
of the unperturbed GR-based Friedmann-Robertson-Walker Universe.  
There are no bumps in it, no 
over-densities, no inhomogeneities, no anisotropies 
and no structure. 
The parameters in it are the background parameters.
It describes the evolution of a perfectly
homogeneous universe \cite{Kelvin}. 

However, bumps are important \cite{Q}.
If there had been no bumps in the CMB thirteen billion 
years ago, no structure would exist today (Fig. 3).
The density bumps seen as the hot and cold spots in the CMB map have 
grown into gravitationally enhanced light-emitting over-densities known 
as galaxies. Their gravitational growth depends on the 
cosmological parameters -- 
much as tree growth depends on soil quality 
(see \cite{E90} for the equations of evolution of the bumps).
Specifically, matching
the power spectrum of the CMB (the $C_{\ell}$ which sample the 
$z \sim 1000$ universe) to the power spectrum of local galaxies ($P(k)$ 
which sample the $z \sim 0$ universe) can be used to constrain 
cosmological parameters.
We measure the bumps and from them we infer the background.

%FRW metric is a simple equation based on isotropy and homogeneity (Lahav, Rees).
%Background is $<\rho>$ with $P(k) = 0$. When there are bumps we have
%$\delta \rho/\rho \neq 0$, $P(k) \neq 0$.
%Bumps can be further divided into linear bumps $\delta \rho/\rho << 1$
%and $\delta T/T << 1$ and non-linear bumps overdensities $\delta \rho/\rho >> 1$.
%(scales smaller than 8 Mpc).
%Bump modes are either adiabatic or isocurvature.
%For the equation governing the growth of the bumps see \cite{E90}.
%The CMB is about as smooth as a cueball.

%\begin{itemize}
%   \item Background, $T_{CMB}$ or $\og$, composition 
%$\om$, $\ol$, $\ob$, $\on$, age:$\t$, expansion $h$.
%   \item Bumps, $n_{s}$,$A_{s}(= Q = C_{10})$, bias parameter $b$, 
%$\Gamma$, $\sigma_{8}$
%\end{itemize}

%%%%%%%%%%%%%%%%%%%%%%%%%%%%%%%%%%%%%%%%%%%%%%%%%%%%
\begin{figure}[p]
    \centering
    \includegraphics[height=11cm,width=8cm,angle=90]{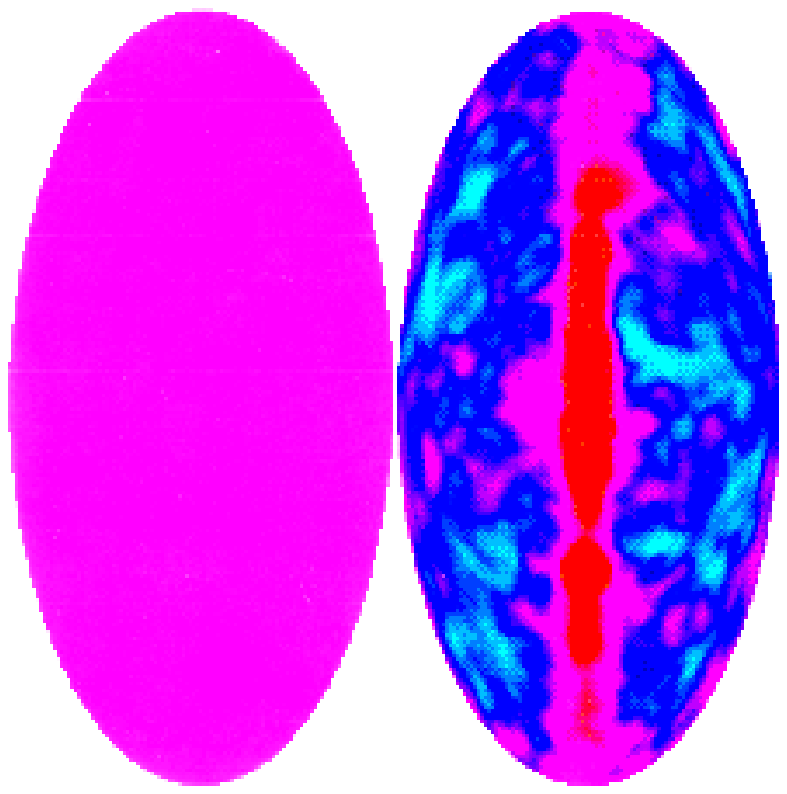}
    \includegraphics[height=7cm,width=5cm]{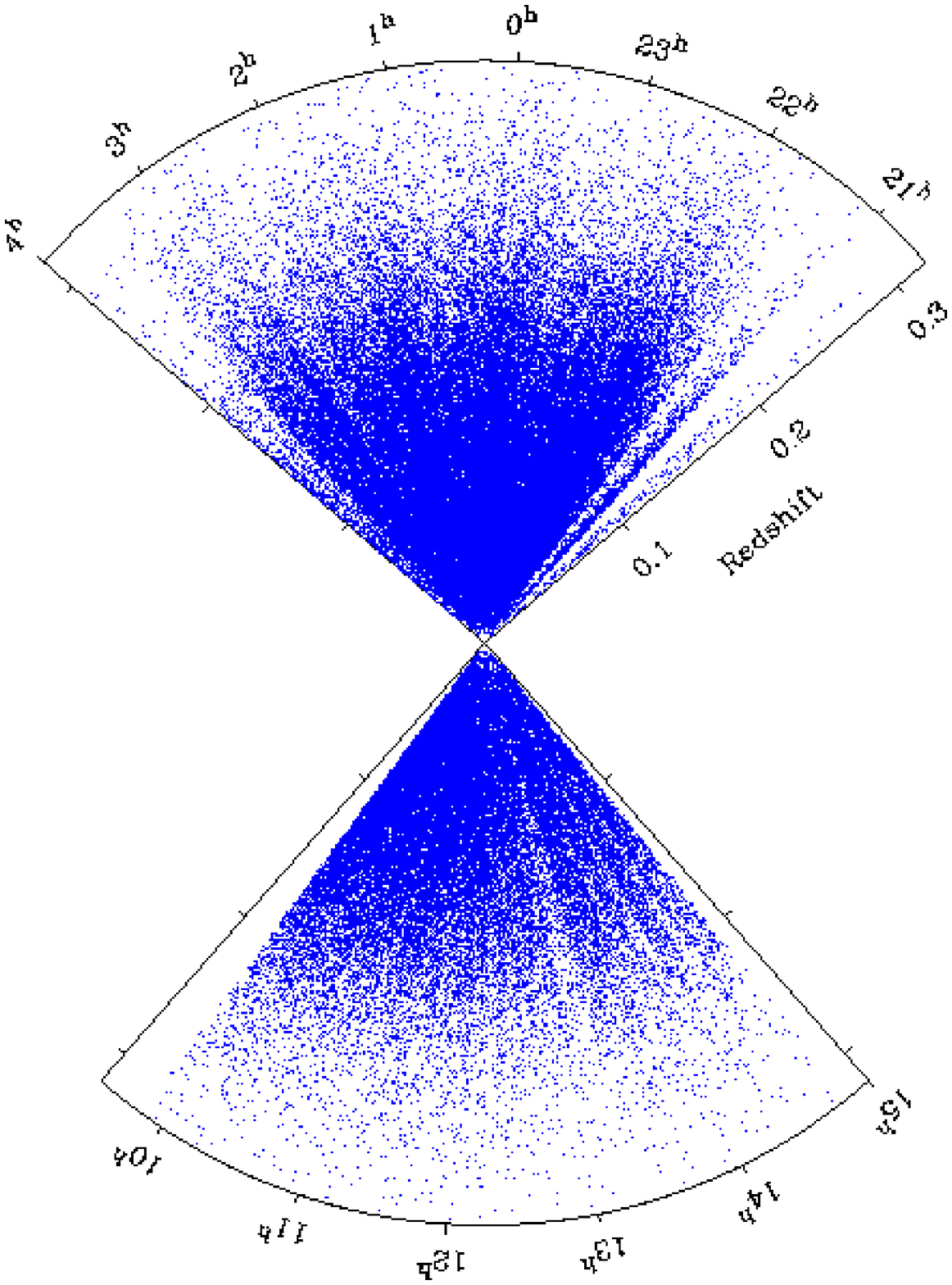}
    \includegraphics[height=7cm,width=5cm]{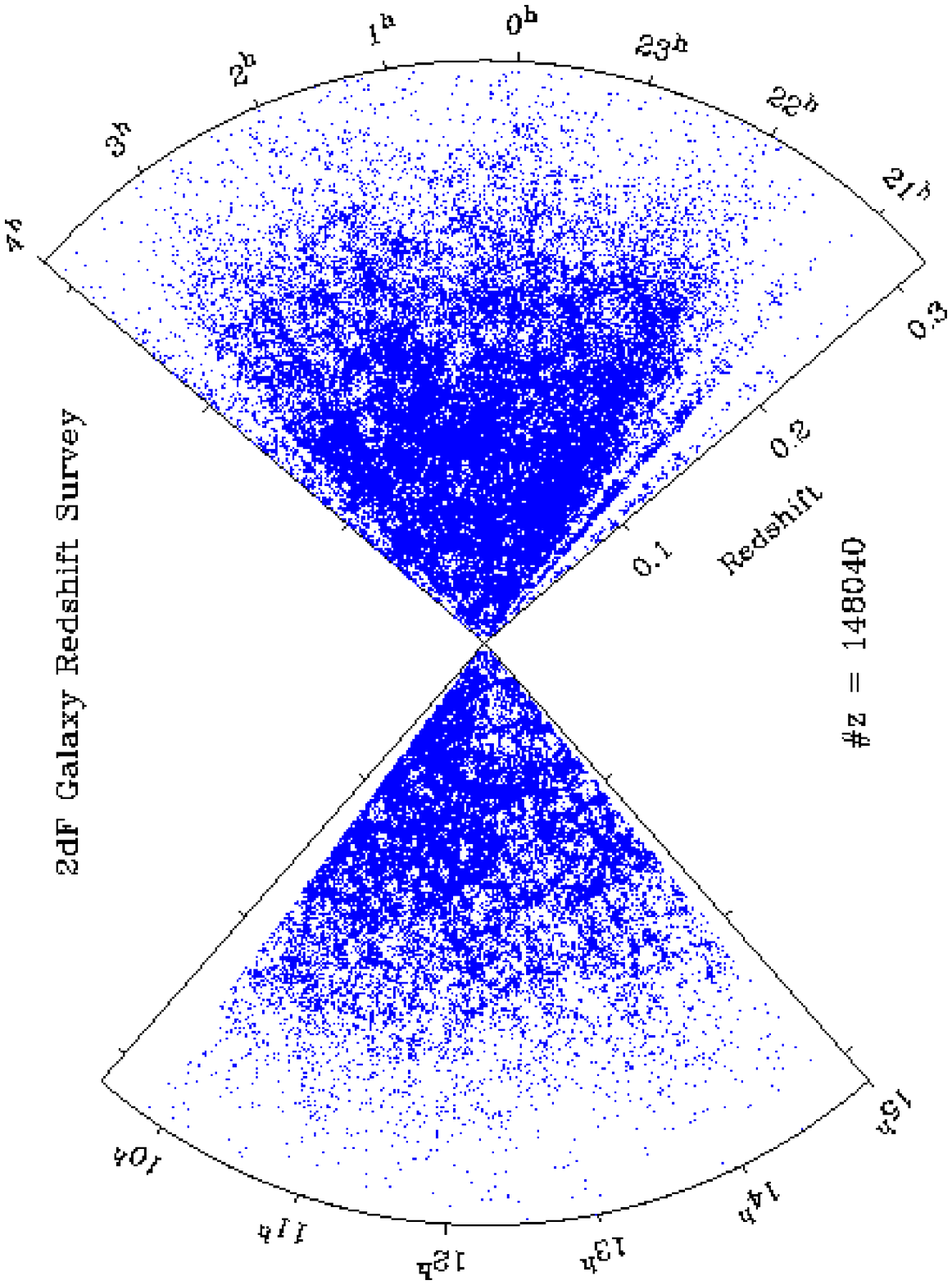}
    \caption{Background and the bumps in it. 
If there were no structure in the Universe then a full-sky microwave 
map would look like the map on the upper left -- perfectly smooth 
and isotropic. The map in the upper right is the COBE-DMR map of the 
CMB at 53 GHz. On a scale of $\pm 150\; \mu$K around the mean 
temperature $2.725$ K, the $\sim 30 \mu$K 
($\delta T/T \sim 10^{-5}$) bumps stand out. The vertical band is our 
Milky Way.  These observable temperature bumps
indicate the presence of density bumps: 
$\frac{\delta \rho}{\rho} = \frac{1}{3}\frac{\delta T}{T}$.
If galaxies were distributed randomly in the Universe with no 
large scale structure, the 2dF galaxy redshift survey of the local universe 
would have produced a map like the one in the lower left.
The map it did produce (lower right) shows galaxies clumped into clusters 
radially smeared by the fingers of God, with 
empty voids surrounded by great walls of galaxies.}
    \label{f:dmr}
\end{figure}

\section{GUTs, TOEs, Branes, Quintessence and Ekpyrotic Cycles}

Many aspects of our Universe are more fundamental, but
harder to estimate, than the cosmological parameters listed above.
For example, the number of dimensions of our Universe is 
a useful parameter that would help us extend the frontiers of 
current research and smooth the inevitable transition from classical to 
quantum cosmology. 
The background and the bump parameters discussed in the previous sections
were classical parameters. 

Models of inflation usually consist of choosing a form for the potential.
Estimates of the slope of the CMB power spectrum $n_{s}$ and its derivative $\frac{dn_{s}}{dk}$ \cite{dndk}
may soon begin to constrain these potentials.
Inflation solves the origin of structure problem with quantum fluctuations,
and is just the beginning of quantum contributions to cosmology.

$\Lambda$CDM is an observational result that has yet to be 
theoretically confirmed. From a quantum field theoretic point of view
$\ol \sim 0.7$ presents a huge problem. It is a quantum term in
a classical equation. But the last time such a 
quantum term appeared in a classical equation, Hawking radiation 
was discovered.
A similar revelation may be in the offing.
The Friedmann equation will eventually be seen as a low energy approximation
to a more complete quantum model in much the same way that 
$\frac{1}{2} m v^{2}$ is a low energy approximation
to $pc$.

% and searches for the short dimensions of a torus
%(simplest non-trivial topology) provide limits but as far as our 
%horizon goes, we seem to be living on a simply connected spacetime
%(Bond, Angelica, Roukema, Luminet references).
%Space is a 3-manifold of nearly constant curvature.

Quantum cosmology is opening up many new doors.
Varying coupling constants are expected at high energy
\cite{Wilczek} and $c$ variation, $G$ variation,
$\alpha$ variation, and $\ol$ variation (quintessence)
are being discussed. 
We may be in an ekpyrotic universe or a cyclic one
\cite{cyclic}.
The topology of the Universe is also not without interest
\cite{Levin}.
Just as we were getting precise estimates of the parameters
of classical cosmology,
whole new sets of quantum cosmological parameters are being proposed
(see Liddle, these proceedings, for a discussion 
of inflation and some of the new cosmological ideas).

But there is hope.
For example, inflationary models and the new ekpyrotic
models 
%\cite{Ekpyrot} 
make different predictions about the slope and amplitude 
of the tensor mode ($n_{T}$ and $A_{T}$) contribution to
the CMB power spectrum.
Measurements of CMB polarization over the next few years 
will add more diagnostic power to CMB parameter estimation
and may be able to usefully constrain $n_{T}$ and $A_{T}$ and 
distinguish these two models.

%Only with the advent of CMBology, precision cosmology, supernovae
%studies, 100K galaxy surveys, observationally probing the earliest epochs of the Universe.
%
%New measurements to watch include
%new CMB observations of anisotropy and polarization from MAP, 
%Planck \cite{MAPPlanck},Boomerang, Viper, DASI, CBI as well as the
%new gravitational wave measurements.

%\section{Summary}
I have presented my version of the current best-fit cosmological parameters.
Other versions %with minor differences 
can be found at \cite{Fukugita}\cite{Scott}\cite{L00} and Turner (this volume).
By estimating cosmological parameters with precision, 
%and with explicit error bars
cosmology has learned to stick its neck out -- it has become a real 
science, error bars and all.
The $\Lambda$CDM model is still the leading candidate.
Our classical picture of the Universe is 
falling into place even as $\ol$ and the new mysteries of quantum cosmology 
enlarge the darkness, keeping the Universe safe for theoretical
rogues and their poorly constrained speculations.

\Acknowledgements
The author is supported by an Australian Research Fellowship and
thanks Matts Roos for a tour through a Finnish swamp looking
for berries in the dusk.

%%%%%%%%%%%%%%%%%%%%%%%%%%%%%%%%%%%%%%%%%%%%%%%%%%%%%%%%%%%%%%%%%%%%%%%%

%%%%%%%%%%%%%%%%%%%%%%%%%%%%%%%%%%%%%%%%%%%%%%%%%%%%%%%%%%%%%%%%%%%
\end{document}